\documentclass{article}

\usepackage{amsfonts}
\usepackage{amsmath}
\usepackage{amssymb}
\usepackage{cite}
\usepackage[dvips]{graphicx}
\usepackage{comment}

\newcommand{\ba}{\begin{eqnarray}}
\newcommand{\ea}{\end{eqnarray}}
\newcommand{\nn}{\nonumber}
\newcommand{\cW}{{\mathcal{W}}}

\newcommand{\be}{\begin{equation}}
\newcommand{\ee}{\end{equation}}

\newcommand{\vY}{\vec{Y}}

\def\b{\beta}

\def\d{\delta}
\def\e{\epsilon}

\def\k{\kappa}

\def\t{\tau}

\def\D{\Delta}

\allowdisplaybreaks[1]

\begin{document}

\begin{titlepage}

\begin{flushright}
UT-13-24\\
TIT/HEP-629
\end{flushright}

\vskip 12mm

\begin{center}
{\LARGE Extended Conformal Symmetry}\\
{\LARGE and}\\
{\LARGE Recursion Formulae for Nekrasov Partition Function}
\vskip 2cm
{\large Shoichi Kanno$^\dagger$\footnote{e-mail: kanno at th.phys.titech.ac.jp}, Yutaka Matsuo$^\flat$\footnote{e-mail: matsuo at phys.s.u-tokyo.ac.jp} and Hong Zhang $^\flat$\footnote{e:mail: kilar at hep-th.phys.s.u-tokyo.ac.jp}}
\vskip 2cm
$^\dagger$ {\it \large Department of Physics, Tokyo Institute of Technology}\\
{\it \large Tokyo, 152-8551, Japan}
\vskip 0.5cm
$^\flat$ {\it \large Department of Physics, The University of Tokyo}\\
{\it \large Tokyo 113-0033, Japan}
\end{center}
\vfill
\begin{abstract}
We derive an infinite set of recursion formulae for Nekrasov instanton
partition function for linear quiver $U(N)$ supersymmetric gauge theories in 4D.  
They have the structure of a deformed 
version of ${\mathcal{W}_{1+\infty}}$ algebra which is called SH$^c$ algebra
(or degenerate double affine Hecke algebra) in the literature. The algebra
contains $W_N$ algebra with general central charge
defined by a parameter $\beta$, which gives the $\Omega$ background
in Nekrasov's analysis.  Some parts of the formulae are identified with the
conformal Ward identity for the conformal block function of Toda field theory.
\end{abstract}
\vfill
\end{titlepage}

\setcounter{footnote}{0}

\newpage
\section{Introduction}

Nekrasov partition function \cite{r:Nekrasov} is an exact formula for the partition function of  four-dimensional $\mathcal{N}=2$
supersymmetric gauge theory, including non-perturbative instanton effects. 
It is calculated in a deformed four-dimensional Euclidean space, called $\Omega$-background,
which is parameterized by two parameters $\epsilon_1,\epsilon_2$.
At the same time, it was recognized that
the partition function can be identified with 
the correlation function of two-dimensional 
conformal field theory. In a recent paper \cite{r:AGT},  
the explicit form of such correspondence was proposed
between $N=2$ gauge theories and Liouville (Toda) conformal blocks
(AGT conjecture).
In AGT proposal, the instanton part of Nekrasov partition function 
is identified with the conformal block of $W$-algebra 
\cite{Wyllard:2009hg, Mironov:2009by} .  

This article is in the line of this development.
The instanton partition function for linear quiver gauge theories
is decomposed into matrix like product with a factor  $Z_{\vec Y, \vec W}$
which depends on two sets of Young diagrams (eq.\ref{e:Nek}).
Here the Young diagrams $\vec Y=(Y_1,\cdots, Y_N)$ 
represent the fixed points of
$U(N)$ instanton moduli space under localization.
$Z_{\vec Y,\vec W}$ consists of contributions from one bifundamental 
hypermultiplet and vectormultiplets.
We find that the building block $Z_{\vec Y, \vec W}$ 
satisfies an infinite series of recursion relations,
\ba \label{e:sketch}
\delta_{\pm 1,n} Z_{\vec Y,\vec W} -U_{\pm 1,n} Z_{\vec Y,\vec W}=0\,,
\ea
where $\delta_{\pm 1,n}Z_{\vec Y,\vec W}$  
represents a sum of the Nekrasov partition function with 
instanton number larger  or less  than $Z_{\vec Y,\vec W}$ by one
with appropriate coefficients and $U_{\pm 1,n}$ are polynomials
of parameters such as the mass of bifundamental matter or the VEV of gauge multilets.
The subscript $n$ takes any non-negative integer values.
The detailed form of the recursion formula  and its derivation 
are given in the first half of this paper.
The recursion formula is derived by a complicated but 
straightforward calculation from the definition of the factor $Z_{\vec Y, \vec W}$.
We note that a classical limit of such relations was
recently explored in \cite{Nekrasov:2012xe}.

In the latter half of this article,  we give an interpretation of (\ref{e:sketch}).
We show that the variation in (\ref{e:sketch}) can be 
understood as an action of an infinite-dimensional extended conformal algebra.
It is defined in \cite{r:SV} and called SH$^c$ algebra.\footnote{This name of
the algebra appears only in \cite{r:SV}. Degenerate
double affine Hecke algebra, or DDAHA in short, may be more appropriate.
We thank Y. Tachikawa for informing us of the relevance of \cite{r:SV}.}
For this purpose, we construct an explicit representation 
where the basis of the Hilbert space is labeled by 
sets of $N$ Young diagrams.
Physically, it can be understood that these states correspond to instantons characterized by the same set of Young diagrams. 
In our previous paper \cite{Kanno:2012hk},  we showed a similar form of recursion formula under self-dual $\Omega$-background $(\e_1+\e_2=0)$
and discussed that it can be interpreted  in terms of $\cW_{1+\infty}$ algebra.
The analysis here is a natural generalization to arbitrary $\Omega$-deformation.
SH$^{c}$ algebra contains  a parameter $\beta$, which is related to 
$\Omega$-deformation parameters by $\beta=-\e_1/\e_2$.
When we take $\beta=1$, (\ref{e:sketch}) reduces to that in \cite{Kanno:2012hk} and 
the action of SH$^c$ algebra can be identified with the  $\cW_{1+\infty}$ algebra.
We will also see SH$^{c}$ algebra contains Heisenberg$\times$Virasoro subalgebra and  
its central charge is the same as that of Heisenberg$\times W_N$ algebra with background charge $Q=\sqrt{\beta}-1/\sqrt{\beta}$. 
The combination of Heisenberg algebra with $W_N$ appears in  \cite{Alba:2010qc,Fateev:2011hq,Belavin:2011js}, where the authors formally construct a basis of Hilbert space of Heisenberg$\times W_N$ algebra which reproduces the
factorized form of Nekrasov partition function. 
Such observation implies that one may regard the formula (\ref{e:sketch})
as the conformal Ward identities which characterize the conformal block function.

We mention that there is another one parameter deformation
of  $\cW_{1+\infty}$ algebra \cite{Gaberdiel:2012ku}, $W_{\infty}[\mu]$
in the context of higher spin supergravity.
SH$^c$ and $W_{\infty}[\mu]$ share a property that they are generated by
infinite higher spin generators and contain
$W_N$ algebra with general $\beta$ as their reduction.
Here we use SH$^c$ since its action on a basis parametrized by
sets of Young diagram is already known.   It is natural to  expect that
these two algebras are identical although they appear to be very different.
It should also be noted that the introduction of further deformation parameter is
possible \cite{r:DAHA, r:DI, r:Miki}
and was applied to a generalization of AGT conjecture \cite{DAHA-AGT}.  

As we will see later, we expect that the recursion relation from SH$^{c}$ algebra
should be regarded as the extended conformal Ward identities and
fully reproduce the conformal block function.
Because of a technical difficulty to characterize the vertex operator
in SH$^c$, explicit demonstration of the relation is limited to
 the Heisenberg and Virasoro subalgebra.          
For these cases, the recursion relations
for $n=0,1$ can be indeed interpreted  as Ward identities. 
The algebra SH$^{c}$ was introduced in \cite{r:SV} to prove the AGT conjecture
for pure super Yang-Mills theory.  Our analysis shows that it may be
applied to linear quiver gauge theories as well.  For the recent development
toward such direction, see also \cite{Maulik:2012wi}.  

The rest of this article is organized as follows. 
In section 2, we describe Nekrasov partition function for linear quiver gauge theories.
In section 3, we derive the recursion formula for Nekrasov partition function. 
In section 4, we give the definition of SH$^c$ algebra  and a  representation of it. 
The relation to $\cW_{1+\infty}$ algebra at $\beta=1$ is also discussed.    
In section 5,  we show  SH$^c$ algebra contains 
Heisenberg$\times$Virasoro subalgebra and its central charge is equal to
that of Heisenberg$\times W_N$ algebra. 
In section 6,  we discuss that Nekrasov partition function can be interpreted as a correlator of SH$^{c}$ algebra.
Especially,  we explain the recursion formulae for $n=0,1$ 
represent the $U(1)$ and Virasoro constraint for Nekrasov partition function respectively. 
The vertex operator in the correlator should be chosen to be special ones which have maximal number
of null states at level 1.
Since the calculations in this article are very lengthy but straightforward, 
most of the detail are not presented.     
We nevertheless keep some outline of the computation
in Appendix for readers who are interested in the detail.  

\section{Nekrasov partition function}
In this article, we consider four-dimensional $\mathcal{N}=2$ superconformal linear quiver gauge theory with 
$U(N) \times U(N) \times \cdots \times U(N)$ gauge group.
The instanton partition function of $N=2$ gauge theories have been
developed in \cite{r:Nekrasov,r:NY,r:MNS,inst1}.
In this case, it can be written in the following form
\ba\label{e:Nek}
Z^\mathrm{Nek}= \sum_{\vec Y^{(1)},\cdots,\vec Y^{(n)}}
q_i^{|\vec Y^{(i)}|} 
\bar{V}_{\vec Y^{(1)}} \cdot Z_{\vec Y^{(1)} \vec Y^{(2)}}\cdots
Z_{\vec Y^{(n-1)}\vec Y^{(n)}} \cdot V_{\vec Y^{(n)}}\,.
\ea
\ba
Z_{\vec Y^{(i)} \vec Y^{(i+1)}}&=& Z(\vec a^{(i)},\vY^{(i)}; \vec a^{(i+1)}, \vY^{(i+1)};\mu^{(i)}),\\
\bar V_{\vY^{(1)}}&=&  Z(\vec\lambda, \vec\emptyset; \vec a^{(1)}, \vY^{(1)};\mu^{(0)}),\\
V_{\vec Y^{(n)}}&=& Z(\vec a^{(n)},\vY^{(n)}; \vec \lambda', \vec\emptyset;\mu^{(n)}),
\ea
where $q_i=\exp{(2\pi i \t_i)}$ represents the complexified coupling constant $\t_i$ of $i$-th $U(N)$ gauge group,  
and $\vec Y^{(i)}$ is a set of $N$ Young diagrams characterizing 
fixed points of localization in the  instanton moduli space of the $i^\mathrm{th}$ $U(N)$.  
$\vec a^{(i)}$ is the VEV for an adjoint scalar field in the vector multiplet of $i^\mathrm{th}$ $U(N)$
and $\mu^{(i)}$ is the mass parameter for the bifundamental matter
field which interpolates $i^\mathrm{th}$ and $i+1^\mathrm{th}$ gauge groups.
We write $\vec\emptyset$ to represent 
a set of null Young diagrams $(\emptyset,\cdots,\emptyset)$. 

\begin{figure}[bpt]
\begin{center}
\includegraphics[scale=0.6]{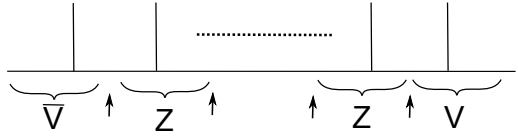}
\end{center}
\caption{Decomposition of Nekrasov function}
\label{f:block}
\end{figure}

The building block reads,
\ba
Z(\vec a, \vec Y; \vec b, \vec W;\mu) =\frac{z_\mathrm{bf}}{z_\mathrm{vect}}= \frac{\prod_{p,q=1}^N g_{Y_p W_q}(a_p-b_q-\mu)}{\left(
\prod_{p,q} g_{Y_p Y_q}(a_p-a_q) g_{W_pW_q}(b_p-b_q)
\right)^{1/2}}
\label{e:Z}
\ea
where the numerator ($z_\mathrm{bf}$) 
comes from the contiribution of the bifundamental multiplet
and the denominator ($z_\mathrm{vect}$) is the contribution from
the vector multiplet to which the bifundamental multiplet couples to. The function $g_{YW}$ is
\ba
g_{Y,W}(x)&=&\prod_{(i,j)\in Y}(x+\beta(Y^\prime_j-i+1)+W_i-j)
\prod_{(i,j)\in W}(-x+\beta(W^\prime_j-i)+Y_i-j+1)\,.
\ea
The decomposition of the form (\ref{e:Nek}) seems to be natural if we recall the pants decomposition of multi-point function on a sphere
and the dictionary of AGT relation; 
A bifundamental and a vector multiplet correspond to a vertex operator insertion and an internal line respectively
(see fig.\ref{f:block}).

\section{Recursion formula for Nekrasov partition function}
\begin{figure}[bpt]
\begin{center}
\includegraphics[scale=0.6]{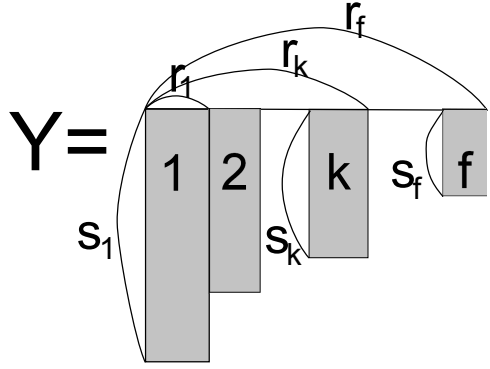}
\end{center}
\caption{Decomposition of Young diagram by rectangles}
\label{f:Young}
\end{figure}

\begin{figure}[bpt]
\begin{center}
\includegraphics[scale=0.6]{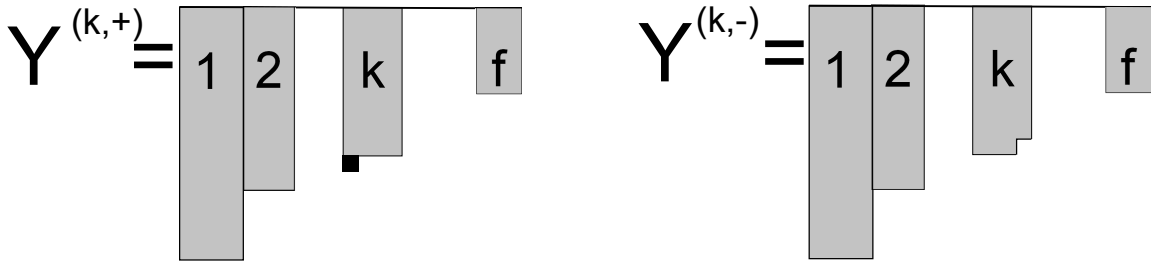}
\end{center}
\caption{Locations of boxes}
\label{f:Young+-}
\end{figure}

In this section, we present the accurate form of the formula (\ref{e:sketch})
and then derive it from the definition (\ref{e:Z}).
%
For this purpose,
we need to introduce some notations.
We decompose $Y,W$ into rectangles $Y=(r_1, \cdots, r_f; s_1,\cdots, s_f)$ 
(with $0<r_1<\cdots <r_f$, $s_1>\cdots>s_f>0$, see Figure \ref{f:Young}
for the parametrization). 
We use $f_p$ (resp. $\bar f_p$) to represent the number of rectangles
of $Y_p$ (resp $W_p$).
Furthermore, we write (with $r_0=s_{k+1}=0$):
\begin{eqnarray}
A_k(Y)& =&  \beta r_{k-1}-s_k-\xi,\quad (k=1,\cdots, f+1)\label{e:Ak}\,,\\
B_k(Y)&= &  \beta r_{k}-s_k,\quad (k=1,\cdots, f)\label{e:Bk}\,,
\end{eqnarray}
where $\xi:=1-\beta$.
$A_k(Y)$ (resp. $B_k(Y)$) represents the $k^\mathrm{th}$ location 
where a box may be added to (resp. deleted from) the Young diagram $Y$  (Figure \ref{f:Young+-})
composed with a map from location to $\mathbf{C}$.
\\\\
We denote $Y^{(k,+)}$ (resp. $Y^{(k,-)}$) as the Young diagram obtained from
$Y$ by adding (resp. deleting) a box at $(r_{k-1}+1,s_k+1)$ (resp. $(r_k,s_k)$).
Similarly we use the notation $\vec Y^{(k\pm),p}=(Y_1,\cdots, Y_p^{(k,\pm)}, \cdots, Y_N)$
to represent the variation of one Young diagram in a set of Young tables $\vec Y$.

One can write the schematic relation (\ref{e:sketch}) more explicitly.
We define,
\ba
\delta_{-1,n} Z(\vec a, \vec Y; \vec b, \vec W;\mu)&=&
\sum_{p=1}^N\left(
\sum_{k=1}^{f_p+1} (a_p+\nu+A_k(Y_p))^n \Lambda^{(k,+)}_p (\vec a,\vec Y) Z(\vec a,\vec Y^{(k,+),p};
\vec b, \vec W;\mu)\right.\nn\\
&& \left. -\sum_{k=1}^{\tilde f_p} (b_p+\mu+\nu+B_k(W_p))^n 
\Lambda^{(k,-)}_p(\vec b, \vec W)
Z(\vec a,\vec Y;\vec b, \vec W^{(k,-),p};\mu)
\right)\label{delta1}\,,\\
\delta_{1,n} Z(\vec a, \vec Y; \vec b, \vec W;\mu)&=&
\sum_{p=1}^N\left(
-\sum_{k=1}^{f_p} (a_p+\nu+B_k(Y_p))^n \Lambda^{(k,-)}_p (\vec a,\vec Y) Z(\vec a,\vec Y^{(k,-),p};
\vec b, \vec W;\mu)\right.\nn\\
&& \left. +\sum_{k=1}^{\tilde f_p} (b_p+\nu+\mu+A_k(W_p)+\xi)^n \Lambda^{(k,+)}_p(\vec b, \vec W)
Z(\vec a,\vec Y;\vec b, \vec W^{(k,+),p};\mu)
\right),\label{delta2}
\ea
where we introduced coefficients $\Lambda$:
\ba
\Lambda^{(k,+)}_p(\vec a,\vec Y) &=& \left(
\prod_{q=1}^N \left(\prod_{\ell=1}^{f_q} \frac{
a_p-a_q+A_k(Y_p)-B_\ell(Y_q)+\xi
}{
a_p-a_q+A_k(Y_p)- B_\ell(Y_q)
}{\prod}_{\ell=1}^{\prime f_q +1}\frac{a_p-a_q+A_k(Y_p)- A_\ell(Y_q) -\xi}{
a_p-a_q+A_k(Y_p)-A_\ell(Y_q)
}
\right)\right)^{1/2}\,,\\
\Lambda^{(k,-)}_p(\vec a,\vec Y) &=& \left(
\prod_{q=1}^N \left(\prod_{\ell=1}^{ f_q+1} \frac{ a_p-a_q+B_k(Y_p)-A_\ell(Y_q)-\xi
}{a_p-a_q+B_k(p)- A_\ell(q)
}{\prod}_{\ell=1}^{\prime  f_q}\frac{a_p-a_q+B_k(Y_p)-B_\ell(Y_q)+\xi}{
 a_p-a_q+B_k(Y_p)-B_\ell(Y_q)
}
\right)\right)^{1/2}\,.
\ea
Prime in the product symbol ($\prod'$) represents that $(\ell,q)=(k,p)$ is excluded
in the product.  The parameter $\nu$ is arbitrary.
\\\\
In order to define the polynomial $U_{\pm 1, n}$, we introduce
a generating function for multi-variables, $x_1,\cdots, x_\mathcal{N},
y_1,\cdots, y_\mathcal{N}$, (the expansion around $\zeta=\infty$),
\ba\label{genfun}
\prod_{I=1}^\mathcal{N} \frac{\zeta-y_I}{\zeta-x_I}=1+\sum_{n=1}^\infty q_{n}(x, y) \zeta^{-n}\,.
\ea
which gives the order $n$ polynomial $q_{n}$ in variables $x_I$ and $y_I$.
$U_{\pm 1,n}$ is written in terms of $q_n$ as
\ba
U_{\pm1,n}=\beta^{-1/2}q_{n+1}(x,y),
\ea
where we need to make replacements of variables:
\ba
x_I &\rightarrow& \{ \nu+A_k(Y_p), \nu+\mu+B_k(W_p)\},\,\,
y_I \rightarrow \{\nu+\mu+A_k(W_p)+\xi , \nu+B_k(Y_p)-\xi\} \quad \mbox{for}\quad U_{-1,n}\,,\label{rep1}\\
x_I &\rightarrow& \{\nu+\mu+A_k(W_p)+\xi , \nu+B_k(Y_p)\},\,\,
y_I\rightarrow \{\nu+A_k(Y_p)+\xi, \nu+\mu+B_k(W_p)\} \quad \mbox{for}\quad U_{1,n}\,.\label{rep2}
\ea
Here $k,p$ run over all possible values and the number of variables
is $\mathcal{N}=N+\sum_{p=1}^N (f_p+\bar f_p)$.

We note that the right hand side of (\ref{genfun}) is written as
\ba
\exp\left(\sum_{n=1}^\mathcal{N} \frac{\zeta^{-n}}{n} p_n(x,y)\right)\,,\quad
p_n(x,y):= \sum_{I=1}^\mathcal{N} ({x_I}^n-{y_I}^n)\,.
\ea
In terms of $p_n$, the function $q_n$ is written as,
\ba
q_1=p_1,\quad q_2=\frac12 (p_2+p_1^2),\cdots
\ea
and so on.  In general it takes the form of Schur polynomial for single row Young diagram $(n)$
written in terms of power sum polynomial.

Let us give a proof of the recursion relation (\ref{e:sketch}).
It is based on a direct evaluation of the variations of
Nekrasov partition function which is given in the appendix \ref{Evaluation of variations of Nekrasov formula}.

By the formulae (\ref{bf1}--\ref{vec4}),
the left hand sides of (\ref{delta1},\ref{delta2}) are written in the form,
\ba
\label{summulti}
\beta^{-1/2}\sum_{I=1}^\mathcal{N} (x_I)^n \frac{\prod_{J=1}^\mathcal{N} (x_I-y_J)}{\prod'_J (x_I-x_J)}
\ea
with the replacements (\ref{rep1},\ref{rep2}).
We rewrite this expression in the form of the generating functional,
\ba
 \sum_{I=1}^\mathcal{N} \left(\sum_{n=0}^\infty
\frac{x_I^n}{\zeta^{n+1}} \right)\frac{\prod_{J=1}^\mathcal{N} (x_I-y_J)}{\prod'_J (x_I-x_J)}
=
\sum_{I=1}^\mathcal{N} \frac{1}{\zeta-x_I} \frac{\prod_{J=1}^\mathcal{N} (x_I-y_J)}{\prod'_J (x_I-x_J)}
= \prod_{I=1}^\mathcal{N} \frac{\zeta-y_I}{\zeta-x_I} -1\,.
\ea
From the second to the third term, we need to use a 
nontrivial identity \cite{Kanno:2012hk} which can be proved 
by comparing the locations of poles and the residue on both hand sides.
The third term takes the form of
the left hand side of (\ref{genfun}).
Comparing the coefficients of $\zeta^{-(n+1)}$, we arrive at
the recursion formula (\ref{e:sketch}).

\section{Symmetry algebra {\bf SH}$^c$}
In this section, we show that the structure of the one box variations in  (\ref{e:sketch})
has a nonlinear algebra which is denoted as SH$^c$ in the paper
\cite{r:SV}.
It has generators $D_{r,s}$  with $r\in \mathbf{Z}$ and $s\in \mathbf{Z}_{\geq 0}$. 
We call the first index $r$ as degree and the second index $s$ as order
of generator.
The commutation relations for degree $\pm 1, 0$ generators are defined by,
\ba
\left[D_{0,l} , D_{1,k} \right] & =& D_{1,l+k-1}, \;\;\; l \geq 1 \,,\label{SH1}\\
\left[D_{0,l},D_{-1,k}\right]&=&-D_{-1,l+k-1}, \;\;\; l \geq 1 \,,\\
\left[D_{-1,k},D_{1,l}\right]&=&E_{k+l} \;\;\; l,k \geq 1\label{eDDE}\,,\\
\left[D_{0,l} , D_{0,k} \right] & =& 0 \,,\,\, k,l\geq 0\,,\label{SH4}
\ea
where 
$E_k$ is a nonlinear combination of $D_{0,k}$ determined in the form of a generating function,
\ba
1+(1-\beta)\sum_{l\geq 0}E_l s^{l+1}= \exp(\sum_{l\geq 0}(-1)^{l+1}c_l \pi_l(s))\exp(\sum_{l\geq 0}D_{0,l+1} \omega_l(s)) \,,\label{com0}
\ea
with 
\ba
&&\pi_l(s)=s^l G_l(1+(1-\beta)s) \,,\\
&&\omega_l(s)=\sum_{q=1,-\beta,\beta-1}s^l(G_l(1-qs)-G_l(1+qs)) \,,\\
&&G_0(s)=-\log(s), \;\;\; G_l(s)=(s^{-l}-1)/l \;\;\; l \geq 1\,.
\ea
The parameters $c_l$ ($l\geq 0$) are central charges.
The first few $E_l$ can be computed more explicitly as,
\ba
E_0&=&c_0,\\
E_1&=&-c_1 +c_0(c_0-1)\xi /2,\label{E_1} \\
E_2&=&c_2+ c_1(1-c_0)\xi +c_0(c_0-1)(c_0-2)\xi^2 /6 +2\beta D_{0,1}, \\
E_3& = &6\beta D_{0,2}+ 2c_0 \beta \xi D_{0,1} + \cdots, \\
E_4
& =& 12\beta D_{0,3}+ 6c_0 \beta \xi D_{0,2} +
(-c_0 \beta \xi^2 +c_0^2 \beta \xi^2- 2c_1 \beta \xi +2 -4\xi +4\xi^2 -2\xi^3)D_{0,1} + \cdots\,.
\ea
where $\cdots$ are terms which does not contain $D_{0,l}$.

Other generators are defined recursively by,
\ba
D_{l+1,0} = \frac1l \left[D_{1,1} , D_{l,0} \right] ,&\qquad&
D_{-l-1,0} = \frac1l \left[D_{-l,0},D_{-1,1}\right] \,, \\
D_{r,l} = \left[D_{0,l+1} , D_{r,0} \right]    \;\;\;  &\qquad&
D_{-r,l}= \left[D_{-r,0} , D_{0,l+1} \right]\,. 
\ea
for $l\geq 0, r>0$\,.

Some of the basic properties of SH$^c$ \cite{r:SV} are listed as follows:
\begin{itemize}
\item The algebra has a natural action on the fixed points of localization in the moduli space of
$SU(N)$ instantons.
\item It can be derived as a singular limit of double affine Hecke algebra (DAHA) \cite{r:DAHA}.
\item When $\beta\rightarrow 1$, the algebra reduces to the much simpler algebra $\cW_{1+\infty}$.
\item For general $\beta$, the algebra contains $W_N$ algebra when the representation is constructed out of $N$ Young diagrams.
\item It is closely related to the recursion relations among Jack polynomials.
\end{itemize}

To see the relation with (\ref{e:sketch}), we introduce a Hilbert space $\mathcal{H}_{\vec a}$
spanned by an basis $|\vec a, \vec Y\rangle$ where $\vec a\in \mathbf{C}^N$
and $\vec Y=(Y_1,\cdots, Y_N)$ is a set of $N$ Young tables.
The dual basis $\langle \vec a, \vec Y|$ is defined such that
\ba
\langle \vec a, \vec Y|\vec b, \vec W\rangle =\delta_{\vec Y, \vec W}\delta(\vec a-\vec b)\,.
\ea
We define the actions of $D_{\pm 1, l}, D_{0,l}$ on the ket and bra basis as,
\ba
D_{-1,l}|\vec b,\vec W>&=&(-1)^{l} \sum_{q=1}^{N} \sum_{t=1}^{\tilde{f_q}}(b_q+B_t(W_q))^l  \Lambda^{(t,-)}_q(\vec W)|\vec b,\vec W^{(t,-),q}>\label{DW1} \,,\\
D_{1,l}|\vec b,\vec W>&=&(-1)^{l}\sum_{q=1}^{N}\sum_{t=1}^{\tilde{f_q}+1}(b_q+A_t(W_q))^l  
\Lambda^{(t,+)}_q(\vec W)|\vec b,\vec W^{(t,+),q}>\label{DW2}\,,\\
D_{0,l+1}|\vec b,\vec W>&=&(-1)^l \sum_{q=1}^{N}\sum_{\mu \in W_q}(b_q+c(\mu))^l |\vec b,\vec W>\,,
\label{eD0W}\\
\langle \vec a, \vec Y| D_{-1,l}&=&(-1)^l\sum_{p=1}^{N}\sum_{t=1}^{f+1}(a_p+A_t(Y_p))^l  \Lambda_p^{(t,+)}(\vec Y)
 \langle \vec a, \vec Y^{(t,+),p}| \,,\label{eDmY}\\
\langle \vec a, \vec Y| D_{1,l}&=& (-1)^l\sum_{p=1}^{N}\sum_{t=1}^{f}
(a_p+B_t(Y_p))^l  {\Lambda}_p^{(t,-)}(\vec Y)\langle \vec a, \vec Y^{(t,-),p}|
\label{eD1Y}\,,\\
\langle \vec a, \vec Y| D_{0,l+1}&=&(-1)^l\sum_{p=1}^{N}\sum_{\mu \in Y_p}(a_p+c(\mu))^l \langle \vec a, \vec Y|\,,\label{eD0Y}
\ea
where
$
c(\mu)=\beta i-j \mbox{ for }\mu=(i,j).   
$

With such definitions, we claim that the action of $D_{a,l}$ 
on the ket and bra basis satisfies SH${}^c$ algebra with central charges
\ba\label{e:cl}
c_l=\left\{
\begin{array}{ll}
\sum_{q=1}^{N}(b_q-\xi)^l \quad &\mbox{(for ket)}\\
\sum_{p=1}^{N}(a_q -\xi)^l\quad &\mbox{(for bra)}
\end{array}
\right. \,.
\ea
We note that the ``central charges" depend on the label $\vec a, \vec b$
in bra and ket state in general except for $c_0 =N$.  Of course, when the inner product between them becomes
nonvanishing ($\vec a=\vec b$), they coincide.

Up to overall signs and shift of parameters $a_p\rightarrow a_p+\nu$
and $b_p\rightarrow b_p+\mu+\nu+\xi$,  the coefficients
which define $D_{\pm 1,l}$ are identical to the variations $\delta_{\pm 1, l}$
in (\ref{delta1},\ref{delta2}).  This observation suggests that
the partition function may be written as an inner product
of the basis $\langle \vec a+\nu\vec e,\vec Y|$ and 
$|\vec b+(\nu+\mu+\xi)\vec e,\vec W\rangle$
 ($\vec e:=(1,\cdots,1)$) with some operator insertions, 
and the recursion formula
should be regarded as the Ward identity for the symmetry algebra SH$^c$.
We will pursue this idea in the following.

Actually there exists a small mismatch in the above observation.
The coefficient appearing in (\ref{delta2}) is shifted from the coefficient
in (\ref{DW1}) by $\xi$.  As we see later, this factor will be canceled
by slightly modifying the vertex operator inserted between two basis.
With such change, the vertex operator is no more the primary field for the $U(1)$ factor.

We need  to perform a lengthy computation to confirm
that the action of $D_{\pm 1, l}$ indeed gives a 
representation of SH$^c$.  See the appendix \ref{s:shc} for some detail.

\subsection{Comparison with $\cW_{1+\infty}$}
For general value of $\beta$, SH$^c$ is a complicated nonlinear algebra.
Simplification occurs when we choose $\beta=1$.
In this case, the nonlinear algebra reduces to a linear algebra
$\cW_{1+\infty}$.   It is an algebra of higher order differential operator $z^n D^m$
($n\in \mathbf{Z}$, $m=0,1,2,\cdots$, $D=z\partial_z$).  Then a quantum generator
$\cW(z^n D^m)$ is assigned to each differential operator (say  $z^n D^m$)
and satisfies the algebra with a central extension,
\ba\label{Winf}
[\cW(z^ne^{xD}) , \cW(z^me^{yD})]=(e^{mx}-e^{ny})\cW(z^{n+m} e^{(x+y)D})-C \frac{e^{mx}-e^{ny}}{e^{x+y}-1}\delta_{n+m,0}
\,.
\ea
The connection between SH$^c$ and $\cW_{1+\infty}$
was already explained in appendix F in \cite{r:SV}.
In our previous paper \cite{Kanno:2012hk},
we use the explicit action of $\cW_{1+\infty}$ generators
on the free fermion Fock space and have shown that
Nekrasov partition function satisfies a recursion formula
associated with the symmetry.

Here we make a direct comparison of the action of $\cW_{1+\infty}$
algebra on the free fermion Fock space in \cite{Kanno:2012hk}
with the corresponding action of SH$^c$ (\ref{DW1}--\ref{eD0W}).
For simplicity, we consider the $N=1$ case. 
\ba
\cW(zD^l)|a,Y\rangle &=& (-1)^l \sum_{i=1}^f (a+B_i(Y)-1)^l |a,Y^{(i,-)}\rangle,\\
\cW(z^{-1}D^l)|a,Y\rangle &=& (-1)^l \sum_{i=1}^{f+1} (a+A_i(Y))^l |a,Y^{(i,+)}\rangle\,.
\ea
We need rewrite $\lambda$ in \cite{Kanno:2012hk} with $-a$ here.
This implies the correspondence in the $\beta\rightarrow 1$ limit:
\ba
D_{-1,l}&\leftrightarrow&  \cW(z (D+1)^l)=  \cW(D^l z),\\
D_{1,l}&\leftrightarrow &\cW(z^{-1} D^l).
\ea
One may proceed to see the correspondence between the generators
in $\cW_{1+\infty}$ and those in SH$^c$.  
The recursion formulae and the Ward identity obtained
in \cite{Kanno:2012hk} can be derived from the corresponding formulae
in this paper by taking the limit $\beta\rightarrow 1$.

\section{Heisenberg  and Virasoro algebra in {\bf SH}$^c$}
In the following, we focus on the important subalgebra in SH$^c$,
namely the Heisenberg (or $U(1)$ current) and Virasoro algebras.
They are important because we can make the explicit evaluation
of Ward identity, while the higher generators in general
have nonlinear commutation relation with the vertex operator.

Generators of Heisenberg ($J_l$) and Virasoro algebras ($L_l$) 
are embedded in SH$^c$ as \cite{r:SV},
\ba
&& J_{l}=(-\sqrt{\beta})^{-l} D_{-l,0},\quad J_{-l}=(-\sqrt{\beta})^{-l} D_{l,0}, \quad J_0=E_1/\beta,
\label{defJ}\\
&& L_l=(-\sqrt{\beta})^{-l} D_{-l,1}/l +(1-l) c_0 \xi J_l/2\,,\quad\nn\\
&& L_{-l}= (-\sqrt{\beta})^{-l} D_{l,1}/l +(1-l)c_0 \xi J_{-l}/2\,,\nn\\
&& L_0=[L_1,L_{-1}]/2=D_{0,1} +\frac{1}{2\beta}\left(c_2+c_1(1-c_0) \xi +\frac{\xi^2}{6} c_0(c_0-1)(c_0-2)\right)\,.\label{defV}
\ea
The commutation relations among these generators are the standard ones,
\ba
\left[ J_n, J_m\right] &=& \frac{n N}{\beta} \delta_{n+m,0},\\
\left[L_n, J_m\right] &=& -m J_{n+m},\\
\left[ L_n, L_m\right]&=& (n-m) L_{n+m}+\frac{c}{12}(n^3-n) \delta_{n+m,0}\,.
\ea
The derivations of these
simple formulae from SH$^c$ commutator are nontrivial
since in the commutation relation of SH$^c$, we have generators with degree$=\pm 1,0$
while $J_n, L_n$ have degree $n$.  
Proof of the first line is given in \cite{r:SV}\,. 
We need derive the commutation relation among 
them recursively.
The confirmation of Virasoro algebra is much more tedious but we give
the explicit computation of $\left[L_2, L_{-2}\right]$ in appendix \ref{derivVir}.
This particular commutation relation is important since
it implies the central charge of Virasoro algebra
is related to those in SH$^c$ as,
\ba\label{cVir}
c= \frac{1}{\beta} \left(
 - c_0^3 \xi^2 +c_0 -c_0 \xi + c_0 \xi^2 
 \right) =1+(N-1) (1-Q^2(N^2+N))\,,\quad
 Q:=\sqrt\beta-\sqrt\beta^{-1}=-\beta^{-1/2}\xi\,.
\ea
This is the central charge for a combined system of $W_N$ algebra
and a free scalar field.  It motivate us to propose a free field representation,
\ba
J(z)&=& \sum_{n} J_n z^{-n-1} =  \beta^{-1/2}\sum_{i=1}^N \partial_z \varphi^{(i)}(z)\,,\\
T(z) &=&  \sum_{n} L_n z^{-n-2} =\sum_{i=1}^N\left(\frac12(\partial\varphi^{(i)}(z))^2 -Q\rho_i \partial^2 \varphi^{(i)}(z)\right)\,, \label{vir}
\ea
with 
\ba
&&\varphi^{(i)} (z)=q^{(i)}+\alpha_0^{(i)} \log z -\sum_{n\neq0} \frac{\alpha^{(i)}_n}{n} z^{-n}\,,\\
&& [\alpha^{(i)}_n, \alpha^{(j)}_m]=n\delta_{n+m,0}\delta_{ij} \,,\quad
[\alpha^{(i)}_m,q^{(j)} ]=\delta_{m,0}\delta_{ij} \,,\\
&& \rho_i=\frac{N+1}{2}-i,\qquad i,j=1,\cdots, N\,.
\ea
Eqs.(\ref{defJ}, \ref{defV}) imply
\ba
J_0|\vec a, \vec Y\rangle&=& \frac{1}{\beta}\left(
-\sum_i (a_i-\xi) +\frac{\xi N(N-1)}{2} \right)|\vec a, \vec Y\rangle,\label{eigenJ}\\
L_0|\vec a, \vec Y\rangle &=&\left( |\vec Y|+\frac{1}{2\beta} 
\left(
\sum_i (a_i-\xi)^2 +(1-N)\xi\sum_i (a_i-\xi) + \frac{\xi^2}{6} N(N-1)(N-2)
\right)\right)|\vec a, \vec Y\rangle\,.\label{eigenL}
\ea
We assign the eigenvalue of $\alpha_0^{(i)}$ on the state $|\vec a, \vec Y\rangle$
as
\ba
\alpha_0^{(i)} |\vec a, \vec Y\rangle\ = p_i|\vec a, \vec Y\rangle\,,
\quad p_i:= -\frac{a_i}{\sqrt{\beta}}-Qi\,,\quad i=1,\cdots, N\,.
\ea
With such assignments, we can rewrite 
(\ref{eigenJ}, \ref{eigenL}) in the more familiar form,
\ba
J_0 |\vec a, \vec Y\rangle=
\frac{1}{\sqrt{\beta}}\left(\vec p\cdot \vec e\right)|\vec a, \vec Y\rangle,\quad
L_0|\vec a, \vec Y\rangle &=& \left(
 |\vec Y| +\Delta(\vec p)\right)|\vec a, \vec Y\rangle,\qquad
 \Delta(\vec p) :=\frac{\vec p\cdot(\vec p-2Q\vec \rho )}{2}\,.
\ea
$\Delta(\vec p)$ is 
the conformal dimension of  a vertex operator $:e^{\vec p \vec\varphi}:$
for (\ref{vir}).

\section{Nekrasov partition function as a correlator and Heisenberg-Virasoro constraints}
In the previous sections, we have seen that the recursion formulae
for Nekrasov partition function takes a form of the representation of
SH$^c$ algebra in terms of the orthonormal basis.  
We have also seen that SH$^c$ algebra contains Heisenberg and 
Virasoro algebras as its subalgebras.  

We observe that AGT conjecture can be proved once we prove the relation
\ba\label{conj}
Z(\vec a, \vec Y; \vec b, \vec W;\mu)=\langle \vec a + \nu \vec e, \vec Y|V(1) |\vec b + (\xi + \nu+\mu) \vec e, \vec W\rangle,
\ea
with the orthonormal basis $|\vec a, \vec Y\rangle$ defined in previous sections
and a vertex operator $V$.  Existence of such basis was formally proved in
\cite{Fateev:2011hq}.
The vertex operator is factorized as $V=\tilde V^H V^W$ where $V^W$ is the
vertex operator for $W_N$ algebra and $\tilde V^H$ describes
the contribution of $U(1)$ factor.  Furthermore it is known that
the correlator of Toda theory is calculable only for the special momenta.
\ba\label{msing}
\vec{p}= -\kappa \vec e_1 \quad \mbox{or}\quad \vec p=-\kappa \vec e_N,
\quad \vec e_1=(1,0,\cdots,0), \quad \vec e_N=(0,\cdots,0,1)\,. 
\ea
The new parameter $\kappa$ is to be determined later.
For the convenience of the computation, we take the latter choice.
$\tilde V^H$ and $V^W$ in the decomposition should be written as,
\ba
\tilde V_\kappa^H=e^{-\frac{\kappa}{N}\vec e\cdot \vec \varphi},\quad
V_\kappa^W=e^{-\kappa(\vec e_N-\frac{\vec e}{N})\vec\varphi}\,,
\label{vmom}
\ea
for $\vec p$ taking the second value in (\ref{msing}).
This form of $W_N$ vertex operator is also important in the context of AGT conjecture.
$V_{\kappa}^W$ is a vertex operator corresponding to the so-called simple puncture. 
As we see, we need modify $\tilde V^H$ to meet the 
behavior of $U(1)$ factor in AGT conjecture.

The relation (\ref{conj}) can be established once one proves that
the partition function $Z$ satisfies the recursion relation which 
defines the right hand side \cite{Kanno:2012hk} .  Namely,
\ba\label{ccond}
0=&&(\langle \vec a + \nu \vec e, \vec Y|D_{n,m}) V(1) |\vec b + (\xi + \nu+\mu) \vec e, \vec W\rangle\nonumber \\
&&-\langle \vec a + \nu \vec e, \vec Y|\left[ D_{n,m}, V(1)\right] |\vec b + (\xi + \nu+\mu) \vec e, \vec W\rangle
-\langle \vec a + \nu \vec e, \vec Y|V(1) (D_{n,m}|\vec b + (\xi + \nu+\mu) \vec e, \vec W\rangle)\,.
\ea
The right hand side gives the Ward identity 
for the conformal block. 
One may translate such relation into a recursion relation
which $Z$ should  satisfy 
if we use the relation (\ref{conj}).
It may sound strange to use the relation to be proved.
Here we use it as the assumption in the inductive method.
It is obvious that the relation (\ref{conj}) holds for the trivial case
$\vec Y=\vec W=\vec\emptyset$
with a proper definition of the inner product.
General relation (\ref{conj}) will be obtained through the Ward identities
by induction.

As we have seen, the recursion relation for $Z$ exists for
$n=\pm 1$ and arbitrary $m\geq 0$.  Other relations should be derived from them.
On the right hand side of \eqref{ccond}, we have already defined
the action of $D_{n,m}$ on the basis.  A problem is that
the commutation relation with the vertex operator cannot be written in 
the closed form except for Heisenberg
and Virasoro generators.  Thus we focus on these cases in the following
though it is not sufficient to complete the inductive proof.


\subsection{Modified vertex operator for $U(1)$ factor}
While the definition of the vertex operator for $W_N$ algebra is
well-known, those for U(1) factor $V^H$ is somewhat tricky
\cite{r:AGT, Fateev:2011hq, r:CO}.\footnote{We thank V. Pasquier to point out this important fact.}
 We give a brief account on the construction.
 
The free boson field which describes the $U(1)$ part is given 
by the operators $J_n$ defined in the previous section.
With
\ba
\alpha_n=\sqrt{\beta/N} J_n,
\ea
we define a free boson field as,
\ba
\phi(z)=q+\alpha_0 \log z -\sum_{n\neq0} \frac{\alpha_n}{n} z^{-n}
=\frac{\vec e \cdot \vec \varphi }{\sqrt{N}}\,.
\ea
We modify the vertex operator $\tilde V^H$ for the $U(1)$ factor as,
\ba
&& V^H_\kappa(z) =e^{\frac{1}{\sqrt N}(NQ-\kappa) \phi_-}
 e^{\frac{-1}{\sqrt N}\kappa \phi_+}\,,\\
&& \phi_+=\alpha_0 \log z -\sum_{n=1}^\infty \frac{\alpha_n}{n} z^{-n}\,,\quad
\phi_-=q +\sum_{n=1}^\infty \frac{\alpha_{-n}}{n} z^{n}\,.
\ea
Such definition of modified vertex operator is needed to reproduce
the contribution of $U(1)$ factor in the correlator \footnote{
Compared with the reference \cite{Fateev:2011hq},
we included the zero mode to modify the commutator with the Virasoro generator.
},
\ba
\langle V^H_{\kappa_1}(z_1)\cdots V^H_{\kappa_n}(z_n)\rangle
=\prod_{i<j}(z_i-z_j)^{\frac{-\kappa_i (NQ-\kappa_j)}{N}}\,.
\ea
Due to the modification, the commutation relation with $U(1)$ current 
(Heisenberg generator) becomes asymmetric,
\ba
\label{u1 commute}
[\alpha_m, V^H_\kappa(z)]=\frac{1}{\sqrt N}(NQ-\kappa) z^m V^H_\kappa(z),\quad
[\alpha_{-n}, V^H_\kappa(z)]=\frac{-1}{\sqrt N}\kappa z^{-n} V^H_\kappa(z)\,,
\ea
for $m\geq0$, $n >0$.

Unlike the standard definition of the 
vertex operator $V=:e^{\kappa \phi}:$, the conformal
property of the modified vertex becomes rather complicated.
It is, however, helpful to understand the recursion relations (\ref{e:sketch})
which has some anomaly as well.
We define the Virasoro generator for the $U(1)$ factor as,
\ba
L^H_n=\frac12 \sum_m :\alpha_{n-m} \alpha_m: \,,
\ea
which has $c=1$.
The commutator of the total Virasoro genrators $L_n=L_n^H+L_n^W$ 
with the vertex $V_\kappa(z)=V^H_\kappa(z) V^W_\kappa(z)$ becomes,
{\small
\ba 
\left[L_n, V_\kappa(z)\right]&=&z^{n+1}\partial_z V_\kappa(z)
+\frac{(NQ-\kappa)^2}{2N}(n+1)z^{n} V_\kappa(z)
+\sqrt{N} Q\sum_{ m =0}^n z^{n-m} V_\kappa(z) \alpha_m
+(n+1)z^{n}\Delta_W V_\kappa(z) ,\,\,\, n\geq 0  \label{LnVp}\,,\\
\left[L_n, V_\kappa(z)\right]&=&z^{n+1}\partial_z V_\kappa(z)
+\frac{\kappa^2}{2N}(n+1)z^{n}  V_\kappa(z)
-\sqrt{N} Q\sum_{ m =1}^{|n|} z^{n+m}\alpha_{-m} V_\kappa(z)
+(n+1)z^{n}\Delta_W V_\kappa(z),\,\,\, n<0 \,,\label{LnVm}
\ea
}
where 
$
\Delta_W=\frac{\kappa(\kappa-Q(N-1))}{2}-\frac{\kappa^2}{2N}
$ is the conformal dimension of $W_N$ vertex operator $V^W_\kappa$ with Toda momenta 
$\vec p=-\kappa(\vec e_N-\frac{\vec e}{N})$ as in (\ref{vmom}).
The anomaly due to the modification of $U(1)$ vertex manifests itself through the third term 
on the right hand side.
We write the commutator for the special cases $n=\pm1, 0$ for the convenience of later calculation.
\ba
\left[L_1, V_\kappa(z)\right]&=&z^{2}\partial_z V_\kappa(z)
+\frac{(NQ-\kappa)^2}{N}z  V_\kappa(z)
+\sqrt{N} Q z V_\kappa(z) \alpha_0
+\sqrt{N} Q  V_\kappa(z) \alpha_1
+2z\Delta_W V_\kappa(z)\,,
\\
 \label{L0 commute}
\left[L_0, V_\kappa(z)\right]&=&z\partial_z V_\kappa(z)
+\frac{(NQ-\kappa)^2}{2N}  V_\kappa(z)
+\sqrt{N} Q V_\kappa(z) \alpha_0
+\Delta_W V_\kappa(z)\,,
\\
\label{L1 commute}
\left[L_{-1},  V_\kappa(z)\right]&=&\partial_z V_\kappa(z) \,.
\ea 
In the following,
we examine the relation (\ref{ccond}) for Heisenberg
($U(1)$) and Virasoro generators for $D_{n,m}$.
\subsection{Ward identities for $U(1)$ currents}
We start from examining the case $n=0$ which can be interpreted as 
the Ward identity for $J_{\pm 1}$,
\ba
&&(\langle \vec a + \nu \vec e, \vec Y|J_{\pm 1}) V(1) |\vec b + (\xi + \nu+\mu) \vec e, \vec W\rangle
-\langle \vec a + \nu \vec e, \vec Y|V(1) (J_{\pm 1}|\vec b + (\xi + \nu+\mu) \vec e, \vec W\rangle) \nn \\
&&=\langle \vec a + \nu \vec e, \vec Y|\left[ J_{\pm 1}, V(1)\right] |\vec b + (\xi + \nu+\mu) \vec e, \vec W\rangle \,.\label{eWJpm}
\ea
By the definition of the representation of  SH$^c$ algebra (\ref{defJ}, \ref{eDmY}, \ref{DW1})
 and the vertex operator (\ref{u1 commute}),
the action of $J_1$ on the bra and ket basis and the commutator with the vertex operator are given as,
\ba
\langle \vec a + \nu \vec e, \vec Y| J_1&=&(-\sqrt{\beta})^{-1}\sum_{p=1}^N\sum_{k=1}^{f_p+1}
\langle \vec a + \nu \vec e,\vec Y^{(k,+),p}| \Lambda^{(k,+)}_p (\vec Y),\\
J_1 |\vec b + (\xi + \nu+\mu) \vec e,\vec W\rangle &=&
(-\sqrt{\beta})^{-1}\sum_{q=1}^N\sum_{\ell=1}^{\tilde f_p} \Lambda^{(\ell,-)}_q (\vec W)|\vec b + (\xi + \nu+\mu) \vec e, \vec W^{(\ell,-),q}\rangle\,, \\
\left[J_1,V_\k(1)\right]&=&\frac{1}{\sqrt \b}(NQ-\kappa) V_\kappa(1)\,.
\ea
Plugging them into (\ref{eWJpm}) gives,
\ba \label{id.J1}
(-\sqrt{\beta})^{-1}\sum_{p=1}^N\sum_{k=1}^{f_p+1} \Lambda^{(k,+)}_p (\vec Y)
\langle \vec a + \nu \vec e,\vec Y^{(k,+),p}| V(1) |\vec b + (\xi + \nu+\mu) \vec e, \vec W\rangle\nonumber  \\
-(-\sqrt{\beta})^{-1}\sum_{q=1}^N\sum_{\ell=1}^{\tilde f_p} \Lambda^{(\ell,-)}_q (\vec W)
\langle \vec a + \nu \vec e, \vec Y|V(1)|\vec b + (\xi + \nu+\mu) \vec e, \vec W^{(\ell,-),q}\rangle\  \\
=\frac{1}{\sqrt \b}(NQ-\kappa)\langle \vec a + \nu \vec e, \vec Y|V(1) |\vec b + (\xi + \nu+\mu) \vec e, \vec W\rangle)\, \nn.
\ea
Using the assumption (\ref{conj}), the left hand side of (\ref{id.J1}) becomes
\ba
\sqrt{\beta}^{-1}\delta_{-1,0} Z(\vec a, \vec Y; \vec b, \vec W;\mu)\,.
\ea
On the other hand, 
taking account of U(1) charge conservation condition, which is derived from the action of $J_0$,
\ba
\label{k requirement}
\kappa=-{\beta}^{-1/2} \sum_{p=1}^N (a_p-b_p-\mu)\,,
\ea
the right hand side of (\ref{id.J1}) becomes 
\ba
\frac{1}{\sqrt \b}(NQ-\kappa)Z(\vec a, \vec Y; \vec b, \vec W;\mu)=\beta^{-1}\sum_{p=1}^{N}(a_p-b_p-\mu-\xi)Z(\vec a, \vec Y; \vec b, \vec W;\mu)
=\sqrt{\beta}^{-1}U_{-1,0}Z(\vec a, \vec Y; \vec b, \vec W;\mu) \,.
\ea
Thus the Ward identity for $J_1$
is proved since it is identified with the recursion formula 
$\delta_{-1,0} Z_{\vec Y,\vec W} -U_{-1,0} Z_{\vec Y,\vec W}=0$.

Derivation of the identity for $J_{-1}$  
can be performed similarly. 
The actions of $J_{-1}$ are given by
\ba
\langle \vec a + \nu \vec e, \vec Y| J_{-1}&=&(-\sqrt{\beta})^{-1}\sum_{p=1}^N\sum_{k=1}^{f_p}
\langle \vec a + \nu \vec e,\vec Y^{(k,-),p}| \Lambda^{(k,-)}_p (\vec Y),\\
J_{-1} |\vec b + (\xi + \nu+\mu) \vec e,\vec W\rangle &=&
(-\sqrt{\beta})^{-1}\sum_{q=1}^N\sum_{\ell=1}^{\tilde f_p+1} \Lambda^{(\ell,+)}_q (\vec W)|\vec b + (\xi + \nu+\mu) \vec e, \vec W^{(\ell+-),q}\rangle\,, \\
\left[J_{-1},V_\k(1)\right]&=&-\frac{1}{\sqrt \b}\kappa V_\kappa(1)\,.
\ea
By the assumption (\ref{conj}), we have
\ba
\label{J-1 difference}
\langle \vec a + \nu \vec e, \vec Y|J_{-1} V_\kappa(1) |\vec b + (\xi + \nu+\mu) \vec e, \vec W\rangle&-&
\langle \vec a + \nu \vec e, \vec Y| V_\kappa(1) J_{-1} |\vec b + (\xi + \nu+\mu) \vec e,\vec W\rangle \nn \\
&=&-\sqrt{\beta}^{-1}\d_{1,0}Z(\vec a, \vec Y; \vec b, \vec W;\mu) \,,
\ea
\ba \label{J-1 com}
\langle \vec a + \nu \vec e, \vec Y|[J_{-1}, V_\kappa(1)] |\vec b + (\xi + \nu+\mu) \vec e,\vec W\rangle=-{\beta}^{-1/2} \kappa 
Z(\vec a, \vec Y; \vec b, \vec W;\mu)\,.
\ea
In the last equality in (\ref{J-1 com}), we use $U(1)$ charge conservation  (\ref{k requirement}).
It shows the equivalence between 
the recursion formula $\delta_{1,0} Z_{\vec Y,\vec W} -U_{1,0} Z_{\vec Y,\vec W}=0$
and the Ward identity for $J_{-1}$. 
We note that the modification of the vertex operator is necessary 
to produce the Ward identities for $U(1)$ currents.


\subsection{Ward identities for Virasoro generators} 
We proceed to examine the equivalence of
the Ward identity for Virasoro generators
and the recursion formula. 
The actions of $L_1$ on the basis and the vertex operator are
evaluated by (\ref{defV}, \ref{DW1}--\ref{eD0Y}, \ref{LnVp}), 
\ba
&&\langle \vec a + \nu \vec e, \vec Y| L_1=\sqrt{\beta}^{-1}\sum_{p=1}^N\sum_{k=1}^{f_p}
\langle \vec a + \nu \vec e, \vec Y^{(k,+),p}|( a_p+ \nu+A_{k}(Y_p))\Lambda^{(k,+),p} (\vec Y), \nn \\
&&L_1 |\vec b + (\xi + \nu+\mu) \vec e
,\vec W\rangle =
\sqrt{\beta}^{-1}\sum_{q=1}^N\sum_{\ell=1}^{f_p} \Lambda^{(\ell,-),q} (\vec W) (b_q+ \nu+\mu+B_{\ell}(W_q)+\xi)|\vec b + (\xi + \nu+\mu) \vec e
,\vec W^{(\ell,-),q}\rangle\,, \nn \\
&&\left[L_1, V_\kappa(1)\right]=\partial V_\kappa(1)
+\frac{(NQ-\kappa)^2}{N} V_\kappa(1)
+\sqrt{N} Q  V_\kappa(1) \alpha_0
+\sqrt{N} Q  V_\kappa(1) \alpha_1
+2\Delta_W V_\kappa(1) \nn
\,.
\ea
As we see from the derivative term in the commutator,
in order to evaluate the Virasoro Ward identities, we need to evaluate 
$\langle\vec a+\nu \vec e,\vec Y|\partial V(1) |\vec b+(\nu+\mu+\xi) \vec e,\vec W\rangle$.
Since the modified vertex operator is not a primary operator, 
the correlator does not have the standard dependence
on the position of the vertex operator. 
We can, however, derive it through the Ward identity of $L_0$.   

According to the actions of $L_0$ on the basis (\ref{eigenL}), we have 
\ba
&&\frac{\langle \vec a + \nu \vec e, \vec Y|L_0 V_\kappa(z) |\vec b + (\xi + \nu+\mu) \vec e
,\vec W\rangle-
\langle \vec a + \nu \vec e, \vec Y| V_\kappa(z) L_0 |\vec b + (\xi + \nu+\mu) \vec e
,\vec W\rangle}{\langle \vec a + \nu \vec e, \vec Y|V_\kappa(z)|\vec b + (\xi + \nu+\mu) \vec e
,\vec W\rangle}\nn\\
&&~~~~~~~~=
\D \left(-\frac{\vec a + \nu \vec e}{\sqrt \b}-Q\vec \rho+Q\frac{N+1}{2}\vec e \right)+|\vec Y|
-\D\left(-\frac{\vec b + (\nu+\mu) \vec e}{\sqrt \b}-Q\vec \rho+Q\frac{N+1}{2}\vec e\right)-|\vec W|\,.
\label{direct L0}
\ea
On the other hand, from the commutator between $L_0$ and vertex operator (\ref{L0 commute}),  we obtain
\ba
&&\left.
\frac{\langle \vec a + \nu \vec e, \vec Y|[L_0,  V_\kappa(z)]|\vec b + (\xi + \nu+\mu) \vec e
,\vec W\rangle}
{\langle \vec a + \nu \vec e, \vec Y|V_\kappa(z)|\vec b + (\xi + \nu+\mu) \vec e
,\vec W\rangle}\right|_{z=1} 
=\left.
\frac{\langle \vec a + \nu \vec e, \vec Y|z\partial_z V_\kappa(1) |\vec b + (\xi + \nu+\mu) \vec e
,\vec W\rangle}{\langle \vec a + \nu \vec e, \vec Y|V_\kappa(1)|\vec b + (\xi + \nu+\mu) \vec e
,\vec W\rangle}\right|_{z=1} \nn \\
&&-\left.\frac{\langle \vec a + \nu \vec e, \vec Y|\sqrt{N} Q V_\kappa(z) \alpha_0|\vec b + (\xi + \nu+\mu) \vec e
,\vec W\rangle}
{\langle \vec a + \nu \vec e, \vec Y|V_\kappa(z)|\vec b + (\xi + \nu+\mu) \vec e
,\vec W\rangle}\right|_{z=1} 
-\frac{(NQ-\kappa)^2}{2N}
-\Delta_W \label{commu L0} \,.
\ea
Since (\ref{direct L0}) is identical with (\ref{commu L0}) by the Ward identity for $L_0$, 
the derivative term can be evaluated as follows,  
\begin{equation}
\begin{split}
\label{partial VZ}
&\frac{\langle \vec a + \nu \vec e, \vec Y|\partial_z V_\kappa(1) |\vec b + (\xi + \nu+\mu) \vec e
,\vec W\rangle}{\langle \vec a + \nu \vec e, \vec Y|V_\kappa(1)|\vec b + (\xi + \nu+\mu) \vec e
,\vec W\rangle}\\ 
=
&\D \left(-\frac{\vec a + \nu \vec e}{\sqrt \b}-Q\vec \rho+Q\frac{N+1}{2}\vec e \right)+|\vec Y|
-\D\left(-\frac{\vec b + (\nu+\mu) \vec e}{\sqrt \b}-Q\vec \rho+Q\frac{N+1}{2}\vec e\right)-|\vec W| \nn \\
&-\frac{\xi}{\beta}\bigg (-\sum_{p=1}^N (b_p +\nu +\mu ) 
+N(N-1)\xi/2\bigg ) 
-\frac{(NQ-\kappa)^2}{2N}
-\frac{\kappa(\kappa-Q(N-1))}{2}+\frac{\kappa^2}{2N} \,.
\end{split}
\end{equation}


Now we are ready to check the recursion relation for Virasoro generators. 
Applying (\ref{conj}), we obtain
\ba \label{L1-1}
&&\langle \vec a + \nu \vec e, \vec Y|L_1 V_\kappa(1) |\vec b + (\xi + \nu+\mu) \vec e
,\vec W\rangle-
\langle \vec a + \nu \vec e, \vec Y| V_\kappa(1) L_1 |\vec b + (\xi + \nu+\mu) \vec e
,\vec W\rangle \nn \\
&&= \sqrt\beta^{-1}\delta_{-1,1} Z(\vec a, \vec Y; \vec b, \vec W;\mu)
-Q\sum_{q=1}^N\sum_{\ell=1}^{f_p} \Lambda^{(\ell,-),q} (\vec W) Z(\vec a, \vec Y; \vec b, \vec W^{(\ell,-),q};\mu) \,.
\ea
Unlike in the $J_1$ case, an additional term appears because the action of SH$^c$ algebra on the ket space is
slightly different from the action of $\d_{-1,1}$ on $Z(\vec a, \vec Y; \vec b, \vec W;\mu)$
as we have explained previously. The commutator part becomes
\ba \label{L1-2}
&&\langle \vec a + \nu \vec e, \vec Y|[L_1, V_\kappa(1)] |\vec b + (\xi + \nu+\mu) \vec e
,\vec W\rangle \nn \\
&&=\bigg\{\D \left(-\frac{\vec a + \nu \vec e}{\sqrt \b}-Q\vec \rho+Q\frac{N+1}{2}\vec e \right)+|\vec Y|
-\D\left(-\frac{\vec b + (\nu+\mu) \vec e}{\sqrt \b}-Q\vec \rho+Q\frac{N+1}{2}\vec e\right)-|\vec W| \nn \\
&&
+\frac{(NQ-\kappa)^2}{2N}
+\frac{\kappa(\kappa-Q(N-1))}{2}-\frac{\kappa^2}{2N} \bigg\} Z(\vec a, \vec Y; \vec b, \vec W;\mu) 
-Q\sum_{q=1}^N\sum_{\ell=1}^{f_p} \Lambda^{(\ell,-),q} (\vec W) Z(\vec a, \vec Y; \vec b, \vec W^{(\ell,-),q};\mu) \nn \\
&&=\sqrt{\beta}^{-1} U_{-1,1} Z(\vec a, \vec Y; \vec b, \vec W;\mu) 
-Q\sum_{q=1}^N\sum_{\ell=1}^{f_p} \Lambda^{(\ell,-),q} (\vec W) Z(\vec a, \vec Y; \vec b, \vec W^{(\ell,-),q};\mu) \,.
\ea
In the last equality we use (\ref{k requirement}). This  also have an anomalous term since the modified vertex is not primary operator 
and its commutator with $L_{1}$ has the $V_\k J_1$ term. 
However, the anomalies in (\ref{L1-1}) and (\ref{L1-2}) are identical and
the Ward identity for $L_1$ is reduced to
the recursion relation $\delta_{-1,1} Z_{\vec Y,\vec W} -U_{-1,1} Z_{\vec Y,\vec W}=0$
which is already proved.
We note that the identity holds only when
we have the special value for the vertex momentum (\ref{msing}).

In the same way, for $L_{-1}$, we have
\ba
&&\langle \vec a + \nu \vec e, \vec Y|L_{-1} V_\kappa(1) |\vec b + (\xi + \nu+\mu) \vec e ,\vec W\rangle-
\langle \vec a + \nu \vec e, \vec Y| V_\kappa(1) L_{-1} |\vec b + (\xi + \nu+\mu) \vec e ,\vec W\rangle \nn \\
&&~~~=\sqrt {\beta}^{-1}\d_{1,1} Z(\vec a, \vec Y; \vec b, \vec W;\mu)\,,  \label{L-1-1}\\
\nn \\
&&\langle \vec a + \nu \vec e, \vec Y|[L_{-1} ,V_\kappa(1)] |\vec b + (\xi + \nu+\mu) \vec e ,\vec W\rangle \nn \\
&&~~~=
\left\{\D \left(-\frac{\vec a + \nu \vec e}{\sqrt \b}-Q\vec \rho+\frac{N+1}{2}\vec e \right)+|\vec Y|
-\D\left(-\frac{\vec b + (\nu+\mu) \vec e}{\sqrt \b}-Q\vec \rho+\frac{N+1}{2}\vec e\right)-|\vec W| \right.\nn \\
&&~~~~~-\frac{\xi}{\beta}\bigg (-\sum_{p=1}^N (b_p +\nu +\mu ) 
+N(N-1)\xi/2\bigg ) 
-\frac{(NQ-\kappa)^2}{2N}
-\frac{\kappa(\kappa-Q(N-1))}{2}+\frac{\kappa^2}{2N} \biggl\} Z(\vec a, \vec Y; \vec b, \vec W;\mu) \nn \\
&&~~=\sqrt{\beta}^{-1} U_{1,1} Z(\vec a, \vec Y; \vec b, \vec W;\mu)\,. \label{L-1-2}
\ea
Again, we use (\ref{k requirement}) to derive the last equality in (\ref{L-1-2}).
Thus, the recursion formula  $\delta_{1,1} Z_{\vec Y,\vec W} -U_{1,1} Z_{\vec Y,\vec W}=0$ can be identified with the Ward identity.
These two consistency conditions are highly nontrivial
and strongly suggest that
the identify (\ref{e:sketch}) are a part of
the Ward identities for the extended conformal symmetry.
\section{Conclusion}
As a generalization of our last work on $\beta =1$ case 
\cite{Kanno:2012hk}, we establish the
recursion relations for arbitrary $\beta$, which characterizes the
Nekrasov partition function and gives a partial proof of AGT conjecture.
This project is much more complicated than before
 and we have to introduce many new ideas to solve the issues caused 
by the arbitrary $\beta$. For example, we have to modify the 
vertex operator to cancel the anomalous terms
in the recursion formulae.
We also need the help of SH$^c$ algebra to define the basis.

Now we have derived the conformal Ward identities for $J_{\pm 1}$ and $L_{\pm 1}$.
The derivation of the similar formulae for the general Virasoro and Heisenberg
generators ($J_n$, $L_n$) will not be so difficult
along the line of \cite{Kanno:2012hk} while the computation
may be tedious and lengthy. What remains to do is to confirm the Ward identities for
$L_{\pm 2}$.  The identities for other generators can be derived from them.
It is supposed to give a proof of
AGT conjecture for $SU(2)$ linear quiver gauge theories.
For the further generalization to $SU(N)$,
we expect that the existence of the recursion formulas for arbitrary $n$ 
in eq.(\ref{e:sketch}) implies
that the Ward identities which completely characterizes the conformal block
may be reduced to  eq.(\ref{e:sketch}) in the end after the proper definition
of the vertex operator in SH$^c$.

We also note 
that there are some important progress in terms of AGT relation \cite{DAHA-AGT}
for the two parameter extension of $\cW_{1+\infty}$ \cite{r:DAHA}.
It is, however, nontrivial to derive AGT from the results of DAHA
since the degeneration limit  is singular.  
We hope to come back to this issue in our future work.

We would like to mention some recent papers which are relevant to this work.
In \cite{r:NSlimit}, 
large $N$ limit ($N$ is the size of Young tableaux) is taken
to relate AGT conjecture to matrix model.  There should be a similar limit
in our recursion formula where the computation becomes much simpler
and the relation with Nekrasov-Shatashvili limit 
\cite{Nekrasov:2009rc} will be clearer.
In \cite{Estienne:2011qk} , the correlator of primary fields is defined in terms of null state condition of $W_N$ algebra 
which in tern relates to Calogero-Sutherland system.  
Since the symmetry of Jack polynomial is identified with SH$^c$, there should be an interesting connection with the current work.
In \cite{Tan:2013tq}, an M-theoretic approach to AGT relation was
explored.
Furthermore, SH$^c$ seems to have interesting applications to 
quantum Hall effects or higher spin 
theories \cite{r:Winf}. These may  also be interesting directions.

\subsection*{Acknowledgments}
We would like to thank  Yuji Tachikawa, Vincent Pasquier, Junichi Shiraishi for their critical comments
at various stages.  We also thank
Didina Serban, Sylvain Ribault and Jean-Emile Bourgine for their interest and comments
in some materials in this paper.
YM is supported in part by KAKENHI (\#25400246). HZ is supported by Global COE Program, the
Physical Sciences Frontier, MEXT, Japan.
SK is supported by JSPS Research Fellowships for Young Scientists.
SK and YM are benefited from the Japan-France exchange 
program SAKURA which enabled the valuable stay at Saclay where this project started.


\appendix
\section{Variations of Nekrasov formula}
\label{Evaluation of variations of Nekrasov formula}

\subsection{A useful formula}

In order to evaluate the variation of Nekrasov partition function
by adding or removing a box, the following formula is essential,
\ba
g_{Y,W}(x)&=&\prod_{(i,j)\in Y}(x+\beta(Y^\prime_j-i+1)+W_i-j)
\prod_{(i,j)\in W}(-x+\beta(W^\prime_j-i)+Y_i-j+1) \label{gyw} \\
&=& P_1 P_2 P_3 Q\,,\nonumber \\
P_1&=& \prod_{(i,j)\in Y} (x+\beta(i-N_2)-j)\,,\qquad
P_2= \prod_{(i,j)\in W} (-x+\beta(i-1)+N_1-j+1)\,,\\
P_3&=& \prod_{i=1}^{N_2} \prod_{j=1}^{N_1} (x+\beta(1-i)-j)^{-1},\qquad
Q= \prod_{i=1}^{N_2} \prod_{j=1}^{N_1} (x+\beta(Y^\prime_j-i+1) +W_i-j)
\ea
where $N_1$ and $N_2$ are arbitrary positive integers which 
should be larger
than the size of Young diagrams $Y,W$.
This  equation is a generalization of Lemma 4 in \cite{Zhang:2011au}.
Since the proof is exactly parallel, we give outline of proof in the following.

We rewrite (\ref{gyw}) in the following form and give a proof,
\begin{equation}
\label{lemma4}
\prod_{i=1}^{N_{2}}\prod_{j=1}^{N_{1}} \frac{x+\b(-i+1)-j }
{x+ \b(Y'_j -i +1) + W_i -j}
=
\prod_{(i,j)\in Y}  \frac{x+\b(i- N_2 ) -j }
{x+ \b(Y'_j -i +1) + W_i -j}
\prod_{(i,j)\in W}  \frac{x+\b(1-i)- N_1+j -1  }
{x -\b(W'_j -i) - Y_i+j-1}\,.
\end{equation}

\paragraph{Proof: }
\underline{\bf  Step 1:} Proof for $W = \emptyset$
is straightforward to do so we omit the details.

\underline{\bf Step 2:} We use the induction in the following. 
Suppose (\ref{lemma4}) is valid for a Young diagram $W$. Let us construct $Z$ which has only one box difference from $W$: $Z_m=W_m +1$, $W'_{W_m +1}=m -1$, $Z'_{W_m +1}=m $, with $m$ the length of $W$.
(Notice that the special case $W_m=0$ means $Z_m$ starts from a new column,
thus we can build any diagram from null Young diagram).
So we just need to prove that
\begin{equation}
\label{lemma4a}
\prod_{i=1}^{N_{2}}\prod_{j=1}^{N_{1}} \frac{x+\b(-i+1)-j }
{x+ \b(Y'_j -i +1) + Z_i -j}
=
\prod_{(i,j)\in Y}  \frac{x+\b(i- N_2 ) -j }
{x+ \b(Y'_j -i +1) + Z_i -j}
\prod_{(i,j)\in Z}  \frac{x+\b(1-i)- N_1+j -1  }
{x -\b(Z'_j -i) - Y_i+j-1}\,.
\end{equation}
The left hand side of (\ref{lemma4a}) is
\begin{equation}
\begin{split}
L
=
\prod_{i=1}^{N_{2}}\prod_{j=1}^{N_{1}}  \frac{x+\b(-i+1)-j }
{x+ \b(Y'_j -i +1) + W_i -j}
\prod_{j=1}^{N_{1}} \frac{x+ \b(Y'_j -m +1) + W_m -j}
{x+ \b(Y'_j -m +1) + W_m +1-j }\;.
\end{split}
\end{equation}
The first factor on the right hand side of (\ref{lemma4a}) is
\begin{equation}
\begin{split}
R_1
=
\prod_{(i,j)\in Y}  \frac{x+\b(i- N_2 ) -j }
{x+ \b(Y'_j -i +1) + W_i -j}
\prod_{j=1}^{Y_m} \frac{x+ \b(Y'_j -m +1) + W_m -j}
{x+ \b(Y'_j -m +1) + W_m+1 -j}\,.
\end{split}
\end{equation}
The second factor becomes
\begin{equation}
\begin{split}
R_2
=
\frac{x+\b(1-m)- N_1+W_m  }
{x - Y_m+W_m}
\times
\prod_{(i,j)\in W}  \frac{x+\b(1-i)- N_1+j -1  }
{x -\b(W'_j -i) - Y_i+j-1}
\prod_{i=1}^{m-1}  \frac{x -\b(m-1 -i) - Y_i+W_m }
{x -\b(m -i) - Y_i+W_m }\;.
\end{split}
\end{equation}
Since we have assumed the equation (\ref{lemma4}) is correct for $W$,
we only need to prove
\begin{equation}
\begin{split}
\label{lemma4b}
\prod_{j=Y_m +1}^{N_1} \frac{x+ \b(Y'_j -m +1) + W_m -j}
{x+ \b(Y'_j -m +1) + W_m+1 -j}
=
\frac{x+\b(1-m)- N_1+W_m  }
{x - Y_m+W_m}
\times
\prod_{i=1}^{m-1}  \frac{x -\b(m-1 -i) - Y_i+W_m }
{x -\b(m -i) - Y_i+W_m }\;.
\end{split}
\end{equation}
The left hand side of the above transforms to
\begin{equation}
\begin{split}
\label{lemma4c}
L'
=
\frac{x+ \b( -m +1) + W_m -N_1}
{x+ \b( -m +1) + W_m -h}
\prod_{j=Y_m +1}^{h} \frac{x+ \b(Y'_j -m +1) + W_m -j}
{x+ \b(Y'_j -m +1) + W_m+1 -j}\;.
\end{split}
\end{equation}
Here $h$ is again the height of $Y$. We name the second term of the last line as $L'_1$\,.
{\small
\begin{equation}
\begin{split}
\label{lemma4d}
L'_1
=
\prod_{j=Y_m +1}^{h}\frac{x+ \b(-m +1) + W_m -j}{x+ \b(-m +1) + W_m+1 -j}
\times
\prod_{j=Y_m +1}^{h}
\prod_{i=1}^{Y'_j}
\bigg(
\frac{x+ \b(i -m +1) + W_m -j}{x+ \b(i-m ) + W_m -j}
\frac{x+ \b(i-m ) + W_m+1 -j}{x+ \b(i -m +1) + W_m+1 -j}
\bigg)\;,
\end{split}
\end{equation}
}
We call the last term of the last line as $L_3$.
The second term on the right hand side of (\ref{lemma4b}) has the form
\begin{equation}
\begin{split}
R'_2
=
\prod_{i=1}^{m-1}
 \frac{x -\b(m-1 -i) +W_m }{x -\b(m -i) +W_m }
\times
\prod_{i=1}^{m-1}
\prod_{j=1}^{Y_i}
\bigg(
\frac{x+ \b(i -m +1) + W_m -j}{x+ \b(i -m +1) + W_m+1 -j}
\frac{x+ \b(i-m ) + W_m+1 -j}{x+ \b(i-m ) + W_m -j}
\bigg)\,,
\end{split}
\end{equation}
so we find 
\begin{equation}
\begin{split}
\label{lemma4e}
\frac{R'_2 }
{L_3}
=
\prod_{i=1}^{m-1}  \frac{x+ \b(i -m +1) + W_m -Y_m}{x+ \b(i -m ) + W_m -Y_m}
=
\frac{x+ W_m -Y_m}{x+ \b(1 -m ) + W_m -Y_m}
\;.
\end{split}
\end{equation}
Combine (\ref{lemma4c}), (\ref{lemma4d}) and (\ref{lemma4e}), it is straightforward to find that
 (\ref{lemma4b}) is tenable, thus complete the proof. \\ \\

\subsection{Variations of Nekrasov formula}
We decompose $Y,W$ into rectangles $Y=(r_1, \cdots, r_f; s_1,\cdots, s_f)$ 
and $W=(t_1,\cdots, t_{\tilde f}; u_1,\cdots, u_{\tilde f})$.
Also we use the same notation such as $Y^{(k,\pm)}$ and $W^{(k,\pm)}$.
For the variation of $Y$ (resp. $W$), $P_2, P_3$ (resp. $P_1,P_3$) remain the same. Variation of $P_1$ 
(resp. $P_2$) produces
a term which cancel the $N_2$ (resp. $N_1$) dependent term in the variation of $Q$.
We also uses a notation $r_0=s_{f+1}=t_0=u_{\tilde f+1}=0$.
After some computation, we obtain,
\ba
\frac{g_{Y^{(k,+)},W}(x)}{g_{Y,W}(x)}&=&\frac{
\prod_{\ell=1}^{\tilde f+1}(x+\beta(r_{k-1}-t_{\ell-1}+1)+u_\ell-s_k-1)
}{
\prod_{\ell=1}^{\tilde f} (x+\beta(r_{k-1}-t_\ell+1) +u_\ell-s_k-1)
}\label{f1}\,,
\\
\frac{g_{Y^{(k,-)},W}(x)}{g_{Y,W}(x)}&=&\frac{
\prod_{\ell=1}^{\tilde f}(x+\beta(r_{k}-t_{\ell})+u_\ell-s_k)
}{
\prod_{\ell=1}^{\tilde f+1} (x+\beta(r_{k}-t_{\ell-1}) +u_\ell-s_k)
}\label{f2}\,,
\\
\frac{g_{Y,W^{(\ell,+)}}(x)}{g_{Y,W}(x)}&=&\frac{
\prod_{k=0}^{f}(-x+\beta(t_{\ell-1}-r_{k})-u_\ell+s_{k+1})
}{
\prod_{k=1}^{f} (-x+\beta(t_{\ell-1}-r_{k}) -u_\ell+s_k)
}\label{f3}\,,
\\
\frac{g_{Y,W^{(\ell,-)}}(x)}{g_{Y,W}(x)}&=&\frac{
\prod_{k=1}^{f}(-x+\beta(t_{\ell}-1-r_{k})-u_\ell+s_{k}+1)
}{
\prod_{k=0}^{f} (-x+\beta(t_{\ell}-r_{k}-1) -u_\ell+s_{k+1}+1)
}\label{f4}\,.
\ea
These expressions becomes more compact by the use of
the notation $A_k(Y_p), B_k(Y_p)$ in (\ref{e:Ak},\ref{e:Bk}),
\ba
\frac{g_{Y_p^{(k,+)} W_q}(a_p-b_q-\mu)}{g_{Y_p W_q}(a_p-b_q-\mu)}
&=& \frac{
\prod_{\ell=1}^{\tilde f_q+1}(a_p-b_q-\mu+A_k(Y_p)-A_\ell (W_q)-\xi)
}{\prod_{\ell=1}^{\tilde f_q}(a_p-b_q-\mu+A_k(Y_p)-B_\ell(W_q))
}\label{bf1}\,,
\\
\frac{g_{Y_p^{(k,-)} W_q}(a_p-b_q-\mu)}{g_{Y_p W_q}(a_p-b_q-\mu)}
&=& \frac{
\prod_{\ell=1}^{\tilde f_q}(a_p-b_q-\mu+B_k(Y_p)-B_\ell(W_q))
}{\prod_{\ell=1}^{\tilde f_q+1}(a_p-b_q-\mu+B_k(Y_p)-A_\ell(W_q)-\xi)
}\,,
\\
\frac{g_{Y_p W_q^{(\ell,+)}}(a_p-b_q-\mu)}{g_{Y_p W_q}(a_p-b_q-\mu)}
&=& \frac{
\prod_{k=1}^{f_p+1}(b_q-a_p+\mu+A_\ell(W_q)-A_k(Y_p))
}{\prod_{k=1}^{f_p}(b_q-a_p+\mu+A_\ell(W_q)-B_k(Y_p)+\xi)
}\,,
\\
\frac{g_{Y_p W_q^{(\ell,-)}}(a_p-b_q-\mu)}{g_{Y_p W_q}(a_p-b_q-\mu)}
&=& \frac{
\prod_{k=1}^{f_p}(b_q-a_p+\mu+B_\ell(W_q)-B_k(Y_p)+\xi)
}{\prod_{k=1}^{f_p+1}(b_q-a_p+\mu+B_\ell(W_q)-A_k(Y_p))
}\label{bf4}\,.
\ea
These are sufficient to calculate variation of $z_\mathrm{bf}$
in (\ref{e:Z}).

To derive the variation of $z_\mathrm{vect}$ for $p\neq q$, we need the 
following formulae which are obtained by putting $W_q\rightarrow Y_q$,
{\small
\ba
\frac{g_{Y_p Y_q}(a_p-a_q)g_{Y_q Y_p}(a_q-a_p)
}{g_{Y^{(k,+)}_p Y_q}(a_p-a_q)g_{Y_q Y_p^{(k,+)}}(a_q-a_p)
}
&=&
\frac{
\prod_{\ell=1}^{f_q}(a_p-a_q+A_k(Y_p)-B_\ell(Y_q) )(a_p-a_q+A_k(Y_p)-B_\ell(Y_q)+\xi)
}{\prod_{\ell=1}^{f_q+1}(a_p-a_q+A_k(Y_p)-A_\ell(Y_q)-\xi)(a_p-a_q+A_k(Y_p)-A_\ell(Y_q))
}\label{vec1}\,,\\
\frac{g_{Y_p Y_q}(a_p-a_q)g_{Y_q Y_p}(a_q-a_p)
}{g_{Y^{(k,-)}_p Y_q}(a_p-a_q)g_{Y_q Y_p^{(k,-)}}(a_q-a_p)
}
&=&
\frac{
\prod_{\ell=1}^{f_q+1}(a_p-a_q+B_k(Y_p)-A_\ell(Y_q)-\xi)(a_p-a_q+B_k(Y_p)-A_\ell(Y_q))
}{\prod_{\ell=1}^{f_q}(a_p-a_q+B_k(Y_p)-B_\ell(Y_q))(a_p-a_q+B_k(Y_p)-B_\ell(Y_q)+\xi)
}\label{vec2}\,.
\ea
}
For the case $p=q$, we obtain,
\ba
\frac{g_{Y_p,Y_p}(0)}{g_{Y_p^{(k+)}Y_p^{(k+)}}(0)}
&=&
\frac{1}{\beta} 
\frac{
\prod_{\ell=1}^{f_q}(A_k(Y_p)-B_\ell(Y_p) )(A_k(Y_p)-B_\ell(Y_p)+\xi)
}{\prod_{\ell=1,(\neq k)}^{f_q+1}(A_k(Y_p)-A_\ell(Y_p)-\xi)(A_k(Y_p)-A_\ell(Y_p))
}\label{vec3}\,,
\\
\frac{g_{Y_p,Y_p}(0)}{g_{Y_p^{(k-)}Y_p^{(k-)}}(0)}
&=&\frac{1}{\beta}
\frac{
\prod_{\ell=1}^{f_p+1}(B_k(Y_p)-A_\ell(Y_p)-\xi)(B_k(Y_p)-A_\ell(Y_p))
}{\prod_{\ell=1}^{f_q}(B_k(Y_p)-B_\ell(Y_p))(B_k(Y_p)-B_\ell(Y_p)+\xi)
}\label{vec4}\,.
\ea
These formulae are sufficient to derive the recursion relation (\ref{e:sketch}).
\section{Derivation of commutation relations of $\bold{SH^c}$ algebra}
\label{s:shc}

First we notice that
\ba
\label{commute}
[D_{-1,k},D_{1,l}] |\vec b,\vec W>&=&(-1)^{k+l}\sum_{q=1}^{N}\left\{ \sum_{t=1}^{\tilde{f}+1}(b_q+A_t(W_q))^{k+l}  
(\Lambda^{(t,+)}_q(\vec b, \vec W))^2\right. \nonumber\\
&& ~~~~ \left.
-\sum_{t=1}^{\tilde{f}}(b_q+B_t(W_q))^{k+l} 
 (\Lambda^{(t,-)}_q(\vec b,\vec W))^2
\right\} |\vec b,\vec W> .
\ea
We have to be careful that the off-diagonal terms, where the
two generator modifies different Young diagrams 
or two different boxes in the same Young diagram,
cancels with each other.  This can be checked as below.
\subsection{Cancellation of  off-diagonal terms}

First, for a Yong diagram with one box removed $W^{(k,-)}$ (or added), we find the relation between $A_t(W^{(k,-)})$, $B_t(W^{(k,-)})$ and their counterparts of the orignal yong diagram $W$.
\ba
\begin{array}{ll}
A_t(W^{(k,-)})=\left\{
\begin{array}{ll}
A_t(W) & 1\leq t \leq k  \\
B_k(W) & t=k+1 \\
A_{t-1}(W) & k+2 \leq t \leq \tilde{f}+2 
\end{array} \right. \,,&
B_t(W^{(k,-)})=\left\{
\begin{array}{ll}
B_t(W) & 1\leq t \leq k-1  \\
B_k(W)-\beta & t=k \\
B_k(W)+1 & t=k+1 \\
B_{t-1}(W) & k+2 \leq t \leq \tilde{f}+1 
\end{array} \right.
\end{array} \,,\nn
\ea

\ba
\begin{array}{ll}
A_s(W^{(k,+)})=\left\{
\begin{array}{ll}
A_s(W) & 1\leq s \leq k-1  \\
A_k(W)-1 & s=k \\
A_k(W)+\beta & s=k+1 \\
A_{s-1}(W) & k+2 \leq s \leq \tilde{f}+2 
\end{array} \right.\,,&
B_s(W^{(k,+)})=\left\{
\begin{array}{ll}
B_s(W) & 1\leq s \leq k-1  \\
A_k(W) & s=k \\
B_{s-1}(W) & k+1 \leq t \leq \tilde{f}+1 
\end{array} \right. 
\end{array} \,.\nn
\ea

With the above relations, we obtain that\footnote{
For simplicity, in the following we do not write $b_q$ explicitly, which always comes together with $A_t( W_q)$ and $B_t( W_q)$.  They may be included by the redefinition of these symbols.},
{\small
\ba
D_{-1,k}D_{1,l}|\vec b,\vec W>&=&\sum_{q=1}^{N} \sum_{t=1}^{\tilde f^{(u,+),\gamma}_q}(B_t(\vec W_q^{(u,+),\gamma}))^k  \Lambda^{(t,-)}_{q}(\vec W^{(u,+),\gamma})
\sum_{\gamma=1}^{N}\sum_{u=1}^{\tilde{f_{\gamma}}+1}(A_{u}(W_{\gamma}))^l 
 {\Lambda}^{(u,+)}_{\gamma}(\vec W)
|\vec b,\vec W^{(u,+),\gamma}_{(t,-),q}>   \,,
\ea
\ba
D_{1,l}D_{-1,k}|\vec b,\vec W>
&=&
\sum_{\gamma=1}^{N}\sum_{u=1}^{\tilde f^{(t,-),q}_{\gamma}+1}
(A_{u}(\vec W_{\gamma}^{(t,-),q}))^l 
 {\Lambda}^{(u,+)}_{\gamma}(\vec W^{(t,-),q})
\sum_{q=1}^{N} \sum_{t=1}^{\tilde{f_q}}(B_{t}(W_k))^k  \Lambda^{(t,-)}_q(\vec W)|\vec b,\vec W^{(t,-),q}_{(u,+),\gamma}>   \,.
\ea
}
For $q = \gamma$, $t \geq u $,
\begin{equation}
\begin{split}
&(A_{\mu}(\vec W_{\gamma}^{(t,-),q}))^l 
\left(
-\prod_{\delta=1}^N \left(\prod_{v=1}^{\tilde f^{(t,-),q}_{\delta}} \frac{
A_{u}(\vec W_{\gamma}^{(t,-),q})-B_{v}(\vec W_{\delta}^{(t,-),q}+\xi
}{
A_{u}(\vec W_{\gamma}^{(t,-),q})- B_{v}(\vec W_{\delta}^{(t,-),q})
}{\prod}_{\delta=1}^{\prime \tilde f^{(t,-),q}_{\delta}+1}\frac{A_{u}(\vec W_{\gamma}^{(t,-),q})- A_{v}(\vec W_{\delta}^{(t,-),q}-\xi}{
A_{u}(\vec W_{\gamma}^{(t,-),q})-A_{v}(\vec W_{\delta}^{(t,-),q}
}
\right)\right)^{1/2}  \\
&= \text{common terms} \times 
\frac{
A_{u}( W_{\gamma}) - (B_{t}( W_{\gamma})-1)
}{
 A_{u}( W_{\gamma})- ( B_{t}( W_{\gamma})-\beta)
}
\times 
\frac{
 A_{u}( W_{\gamma})- (B_{t}( W_{\gamma})+\beta)
}{
 A_{u}( W_{\gamma}) - ( B_{t}( W_{\gamma})+1)
}
\times 
\frac{
 A_{u}( W_{\gamma}) - ( B_{t}( W_{\gamma})+\xi)
}{
 A_{u}( W_{\gamma}) -B_{t}( W_{\gamma})
} \,,
\end{split}
\end{equation}
{\small
\begin{equation}
\begin{split}
&(B_{t+1}(\vec W_q^{(u,+),\gamma}))^k 
 \left(
-\prod_{p=1}^N \left(\prod_{s=1}^{\tilde f^{(u,+),\gamma}_p+1} \frac{
 B_{t+1}(\vec W_q^{(u,+),\gamma})- A_{s}(\vec W_p^{(u,+),\gamma})-\xi
}{
 B_{t+1}(\vec W_q^{(u,+),\gamma})- A_{s}(\vec W_p^{(u,+),\gamma})
}{\prod}_{s=1}^{\prime \tilde f^{(u,+),\gamma}_p}\frac{ B_{t+1}(\vec W_q^{(u,+),\gamma})-B_{s}(\vec W_p^{(u,+),\gamma})+\xi}{
 B_{t+1}(\vec W_q^{(u,+),\gamma})-B_{s}(\vec W_p^{(u,+),\gamma})
}
\right)\right)^{1/2} \\
&= \text{common terms} \times 
\frac{
 B_{t}(W_q)-(A_{u}(W_q)-\beta)
}{
 B_{t}(W_q)-  (A_{u}(W_q)-1)
}
\times 
\frac{
 B_{t}(W_q)-(A_{u}(W_q)+ 1)
}{
 B_{t}(W_q)-  (A_{u}(W_q)+ \beta)
} 
\times 
\frac{
 B_{t}(W_q)-(A_{u}(W_q)-\xi)
}{
 B_{t}(W_q)-  A_{u}(W_q)
}\,.
\end{split}
\end{equation}
}
Thus we find that $\sum^{(t,-),\gamma}_{(\gamma),u}$ cancels with $\sum^{(u,+),\gamma}_{(\gamma),t+1}$. For $q = \gamma$, $t \leq u -2$, we have the same result. For $q = \gamma$, $t = u -1$, we have the direct sum. For $q \neq \gamma$, using a similar method, we also find
that $\sum_q\sum_{\gamma}$ cancels with $\sum_{\gamma}\sum_q$.\\
In total we show that all the off-diagonal terms are gone.

\subsection{Evaluation of  diagonal terms}
Since the right hand side  of \eqref{commute} only depends on $k+l$, we have $ [D_{-1,k},D_{1,l}]=[D_{-1,0},D_{1,l+k}]$.
We need to define the action of $D_{0,l}$. For this purpose, we consider   
\ba
X(s)=<\vec b ,\vec W|\sum_{l \geq 0}[D_{-1,0},D_{1,l}]s^l |\vec b,\vec W>.
\ea
Then from the definition of algebra, we obtain
{\small
\begin{equation}
\begin{split}
&s\xi X(s) =\sum_{q=1}^{N}\left\{
\sum_{t=1}^{\tilde f +1}\frac{s\xi}{1+s (b_q+A_t(W_q))}(\Lambda^{(t,+)}_q(\vec b,\vec W))^2
- \sum_{t=1}^{\tilde f}\frac{s\xi}{1+s(b_q+B_t(W_q))}(\Lambda^{(t,-)}_q(\vec b,\vec W))^2
\right\}  \\
&=\sum_{q=1}^{N}\bigg\{ 
\sum_{t=1}^{\tilde {f}+1}\frac{s\xi}{1+s (b_q+A_t(W_q))}\prod_{p=1}^{N}\big\{\prod_{k=1}^{\tilde{f}}
\frac{b_q-b_p+A_t(W_q)-B_k(W_p)+\xi}{b_q-b_p+A_t(W_q)-B_k(W_p)} \prod_{k \neq t}^{\tilde{f}+1}\frac{b_q-b_p+A_t(W_q)-A_k(W_p)-\xi}
{b_q-b_p+A_t(W_q)-A_k(W_p)}\big\}
 \\
&  -\sum_{t=1}^{\tilde f}\frac{s\xi}{1+s(b_q+B_t(W_q))}\prod_{p=1}^{N}\big\{\prod_{(p),k \neq (q),t}^{\tilde{f}}\frac{b_q-b_p+B_t(W_q)-B_k(W_p)+\xi}{b_q-b_p+B_t(W_q)-B_k(W_p)} \prod_{k=1}^{\tilde{f}+1}\frac{b_q-b_p+B_t(W_q)-A_k(W_p)-\xi}{b_q-b_p+B_t(W_q)-A_k(W_p)}\big\}
\bigg\}  \\
&= -1+\prod_{q=1}^{N}\prod_{t=1}^{\tilde{f}}\frac{1+s(b_q+B_t(W_q)-\xi)}{1+s(b_q+B_t(W_q))}
\prod_{t=1}^{\tilde{f}+1}\frac{1+s (b_q+A_t(W_q)+\xi)}{1+s (b_q+A_t(W_q))}. \label{com1}
\end{split}
\end{equation}
}
The last equality holds because the both sides (i) are degree 0 rational function in $s$, (ii) have the same simple poles and residues at $s=-1/{(b_q+B_t(W_q))}, -1/{(b_q+A_t(W_q))}$ and (iii) vanish at $s=0$.   
We can rewrite (\ref{com1}) as
{\small
\ba
&&1+s\xi X(s)=\prod_{q=1}^{N}\prod_{t=1}^{\tilde{f}}\frac{1+s(b_q+B_t(W_q)-\xi)}
{1+s(b_q+B_t(W_q))}\prod_{t=1}^{\tilde{f}+1}\frac{1+s (b_q+A_t(W_q)+\xi)}{1+s (b_q+A_t(W_q))} \nn \\
&=&\exp \left\{ \sum_{q=1}^{N}\sum_{l=1}^{\infty} \frac{(-1)^ls^l}{l} (\sum_{t=1}^{\tilde{f}}
(p_l(b_q+B_t(W_q))-p_l(b_q+B_t(W_q)-\xi)+
\sum_{t=1}^{\tilde{f+1}}(p_l(b_q+A_t(W_q))-p_l(b_q+A_t(W_q)+\xi))\right\} \,,\nn \\ 
\ea
}
where $p_l(x_I)=\sum_{I} x^l$. 

We define $H_l(W_q):=\sum_{t=1}^{\tilde{f}}(p_l(b_q+B_t(W_q))-p_l(b_q+B_t(W_q)-\xi)+
\sum_{t=1}^{\tilde{f}+1}(p_l(b_q+A_t(W_q))-p_l(b_q+A_t(W_q)+\xi))$.
Then we use a formula,
\ba
H_l(W_q)&=&(b_q-\xi)^l-(b_q)^l-\sum_{\mu \in W_q} \sigma_l(c_q(\mu)) \,,
\ea
where $\sigma_l(x)=(x+1)^l-(x-1)^l+(x-\beta)^l-(x+\beta)^l+(x+\beta-1)^l-(x+1-\beta)^l $ and
$c_q(\mu)=b_q+\beta i-j $ for $\mu=(i,j)$. 
It was proved in appendix B of \cite{r:SV}.
Thus we can proceed as
\ba
1+\xi s X(s)&=&\exp\left\{\sum_{q=1}^{N}\sum_{l=1}^{\infty}(-1)^l\frac{s^l}{l}((b_q-\xi)^l-(b_q)^l)-\sum_{q=1}^{N}\sum_{\mu \in W_q}\sum_{l=1}^{\infty}(-1)^l\frac{s^l}{l}\sigma_l((c_q(\mu))\right\} \nn \\
&=&\exp\left\{\sum_{q=1}^{N}\sum_{l=0}^{\infty}(-1)^{l+1}(b_q-\xi)^l \pi_l(s)\right\} \exp\left\{\sum_{q=1}^{N}{\sum_{l=0}^{\infty} (\sum_{\mu \in W_q}(-1)^l c_q(\mu)^l) \omega_l(s)}\right\} \label{com2} \,.
\ea
In the last equality of (\ref{com2}), we use the following formula
\ba
\sum_{l=1}^{\infty}(-1)^{l+1}\frac{s^l}{l}\{(a+b)^l-a^l \}=\sum_{l=0}^{\infty}(-1)^{l+1}a^ls^lG_l(1+bs)\,, \label{fomu}
\ea
 which can be proved using the taylor expansion of $\rm log (1+s(a+b))$           .
Comparing (\ref{com2}) with (\ref{com0}), the algebra (\ref{eDDE}) is proved
once we set (\ref{eD0W}, \ref{e:cl}).
The proof of the algebra for the action on the bra state is similar.

\section{Derivation of Virasoro algebra from SH$^c$} \label{derivVir}
Here we give a sample computation to give the Virasoro algebra
from the definition of SH$^c$ (\ref{SH1}--\ref{SH4}) and (\ref{defV}).
We focus on the relation
\ba
[L_2,L_{-2}]=4 L_0 +\frac{c}{2} 
\ea
since it gives the simplest commutator to give the Virasoro central charge.

The definition of generators gives
\ba
\left[L_2 , L_{-2} \right]  & =& \frac{1}{4\beta^2}\left\{[D_{-2,1} , D_{2,1}]
-c_0 \xi [D_{-2,1} , D_{2,0}]-c_0 \xi [D_{-2,0} , D_{2,1}]
 +(c_0 \xi)^2 [D_{-2,0} , D_{2,0}]
 \right\} \,.\nn 
\ea
We express degree 2 generator as the commutator of
degree 1 generator
\ba
D_{2,0}    =  \left[D_{1,1} , D_{1,0} \right],&\quad&
D_{-2,0}  = \left[D_{-1,0} , D_{-1,1} \right],\nn \\
D_{2,1} =\left[D_{1,2} , D_{1,0} \right], &\quad &
D_{-2,1}  = \left[D_{-1,0} , D_{-1,2} \right] .
\ea
The commutation relation between degree two operators is reduced to
those for degree one operator.  After some computation we arrive at
\ba
\left[D_{-2,1} , D_{2,1} \right] & = & 8\beta E_2 + 6c_0 \beta \xi E_1 -c_0^2 \beta \xi^2
 +c_0^3 \beta \xi^2 - 2c_0 c_1 \beta \xi +2c_0\beta -2c_0\beta\xi +2c_0\beta\xi^2 \,, \\ 
\left[D_{-2,0} , D_{2,1} \right]  &=& -4c_1\beta +4 c_0^2 \beta \xi -2 c_0 \beta \xi \,, \\
\left[D_{-2,1} , D_{2,0} \right]  
&=&  -4c_1\beta +4 c_0^2 \beta \xi -2 c_0 \beta \xi \,,\\
\left[D_{-2,0} , D_{2,0} \right]  
&=&  2 c_0 \beta \,.
\ea
It gives
\ba
\left[L_2 , L_{-2} \right]  & =&\frac{4}{2\beta} E_2 +
 \frac1{2\beta}  \left\{
 - c_0^3 \xi^2 +c_0 -c_0 \xi + c_0 \xi^2 
 \right\} \,.
\ea
After identifying $L_0=\frac{1}{2\beta}E_2$, we can identify
the Virasoro central charge (\ref{cVir}).

\end{document}